\documentclass[twocolumn,floatfix,showpacs,superscriptaddress]{revtex4}
\usepackage{graphicx}
\usepackage{amsmath}
\usepackage{bm}
\begin{document}

\title{How spin-orbit interaction can cause electronic shot noise}
\author{A. Ossipov}
\affiliation{Instituut-Lorentz, Universiteit Leiden, P.O. Box 9506, 2300 RA Leiden, The Netherlands}
\author{J. H. Bardarson}
\affiliation{Instituut-Lorentz, Universiteit Leiden, P.O. Box 9506, 2300 RA Leiden, The Netherlands}
\author{C. W. J. Beenakker}
\affiliation{Instituut-Lorentz, Universiteit Leiden, P.O. Box 9506, 2300 RA Leiden, The Netherlands}
\author{J. Tworzyd{\l}o}
\affiliation{Institute of Theoretical Physics, Warsaw University, Ho\.{z}a 69, 00--681 Warsaw, Poland}
\author{M. Titov}
\affiliation{Department of Physics, Konstanz University, D--78457 Konstanz, Germany}
\date{March 2006}
\begin{abstract}
The shot noise in the electrical current through a ballistic chaotic quantum dot with $N$-channel point contacts is suppressed for $N\rightarrow\infty$, because of the transition from stochastic scattering of quantum wave packets to deterministic dynamics of classical trajectories. The dynamics of the electron spin remains quantum mechanical in this transition, and can affect the electrical current via spin-orbit interaction. We explain  how the role of the channel number $N$ in determining the shot noise is taken over by the ratio $l_{\rm so}/\lambda_{F}$ of spin precession length $l_{\rm so}$ and Fermi wave length $\lambda_{F}$, and present computer simulations in a two-dimensional billiard geometry (Lyapunov exponent $\alpha$,  mean dwell time $\tau_{\rm dwell}$, point contact width $W$) to demonstrate the scaling $\propto(\lambda_{F}/l_{\rm so})^{1/\alpha\tau_{\rm dwell}}$ of the shot noise in the regime $\lambda_{F}\ll l_{\rm so}\ll W$.
\end{abstract}
\maketitle

Electrical conduction is not much affected typically by the presence or absence of spin-orbit interaction. A familiar example \cite{Ber84,Bro02,Zum02,Zai05}, the crossover from weak localization to weak anti-localization with increasing spin-orbit interaction, amounts to a relatively small correction to the classical conductance, of the order of the conductance quantum $e^{2}/h$. The relative smallness reflects the fact that the spin-orbit interaction energy $E_{\rm so}$ is much smaller than the Fermi energy $E_{F}$, basically because $E_{\rm so}$ is a relativistic correction to the kinetic energy \cite{Win03}.

In this paper we identify an effect of spin-orbit interaction on the electrical current that has a quantum mechanical origin (like weak anti-localization), but which is an order-of-magnitude effect rather than a correction. The effect is the appearance of shot noise in a ballistic chaotic quantum dot with a large number $N$ of modes in the point contacts. According to recent theory \cite{Aga00,Sil03} and experiment \cite{Obe02}, the shot noise without spin-orbit interaction is suppressed exponentially $\propto\exp(-\tau_{E}/\tau_{\rm dwell})$ when the Ehrenfest time $\tau_{E}\simeq\alpha^{-1}\ln N$ becomes greater than the mean dwell time $\tau_{\rm dwell}$ of an electron in the quantum dot. (The coefficient $\alpha$ is the Lyapunov exponent of the classical chaotic dynamics.) The suppression occurs because electrons follow classical deterministic trajectories up to $\tau_{E}$ (in accord with Ehrenfest's theorem, hence the name ``Ehrenfest time''). If $\tau_{E}>\tau_{\rm dwell}$ an electron wave packet entering the quantum dot is either fully transmitted or fully reflected, so no shot noise appears \cite{Bee91}.

\begin{figure}
\centerline{\includegraphics[width=8cm]{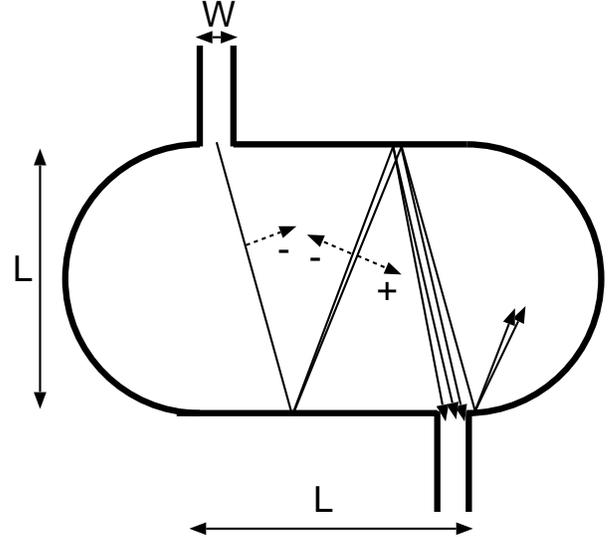}}
\caption{
Splitting of trajectories by spin-orbit interaction in an electron billiard. (The dotted arrows indicate the spin bands, with $\pm$ helicities.) The splitting produces shot noise if not all trajectories can exit through the same opening. 
}
\label{fig_mechanism}
\end{figure}

The electron spin of $\pm\frac{1}{2}\hbar$ remains quantum mechanical in the limit $N\rightarrow\infty$. In the presence of spin-orbit interaction the quantum mechanical uncertainty in the spin of the electron is transferred to the position, causing a breakdown of the deterministic classical dynamics and hence causing shot noise. The mechanism for the spin-orbit-interaction-induced shot noise is illustrated in Fig.\ \ref{fig_mechanism}. The key ingredient is the splitting of a trajectory upon reflection with a hard boundary \cite{Gov04}.

Whether a boundary is ``hard'' or ``soft'' depends on the relative magnitude of the penetration depth $\xi$ into the boundary and the spin-orbit precession length $l_{\rm so}=h v_{F}/E_{\rm so}\simeq \lambda_{F} E_{F}/E_{\rm so}$. A soft boundary has $\xi\gg l_{\rm so}$, so the spin evolves adiabatically during the reflection process \cite{Sil05} and the electron remains in the same spin band, without splitting of the trajectory. In the opposite regime $\xi\ll l_{\rm so}$ of a hard boundary the spin is scattered into the two spin bands by the reflection process. The energy splitting $E_{\rm so}$ of the spin bands at the Fermi level amounts to a difference $\delta p_{\perp}\simeq E_{\rm so}/v_{F}$ of the component of the momentum perpendicular to the boundary, and hence to a splitting of the trajectories by an angle $\delta\phi_{\rm so}\simeq\delta p_{\perp}/p_{F}\simeq \lambda_{F}/l_{\rm so}$. (A precise calculation of the splitting, which depends on the angle of incidence, will be given later.)

Because of the chaotic dynamics, the angular opening $\delta\phi_{\rm so}(t)\simeq(\lambda_{F}/l_{\rm so})e^{\alpha t}$ of a pair of split trajectories increases exponentially with time $t$ --- until they leave the dot through one of the two point contacts after a time $T$. The splitting will not prevent the trajectories to exit together through the same point contact if $\delta\phi_{\rm so}(T)\lesssim W/L$, with $W$ the width of the point contact and $L$ the diameter of the (two-dimensional) quantum dot. The time
\begin{equation}
T_{\rm so}=\alpha^{-1}\ln(W l_{\rm so}/L\lambda_{F}) \label{Tsodef}
\end{equation}
at which $\delta\phi_{\rm so}(T_{\rm so})=W/L$ is an upper bound for deterministic noiseless dynamics due to spin-orbit scattering. 

Dwell times shorter than $T_{\rm so}$ may yet contribute to the shot noise as a result of diffraction at the point contact, which introduces an angular spread $\delta\phi_{\rm pc}\simeq 1/N\simeq \lambda_{F}/W$ in the scattering states. The time
\begin{equation}
T_{\rm pc}=\alpha^{-1}\ln(WN/L) \label{Tpcdef}
\end{equation}
at which this angular spread has expanded to $W/L$ is an upper bound for deterministic noiseless dynamics due to diffraction at the point contact \cite{Sil03}. The smallest of the two times $T_{\rm so}$ and $T_{\rm pc}$ is the Ehrenfest time of this problem,
\begin{equation}
\tau_{E}=\alpha^{-1}\ln\bigl[(W/L)\min(N,l_{\rm so}/\lambda_{F})\bigr],\label{tauEdef}
\end{equation}
separating deterministic noiseless dynamics from stochastic noisy dynamics. (By definition, $\tau_{E}\equiv 0$ if the argument of the logarithm is $<1$.) Since the distribution of dwell times $P(T)\propto\exp(-T/\tau_{\rm dwell})$ is exponential, a fraction $\int_{\tau_{E}}^{\infty}P(T)\,dt=\exp(-\tau_{E}/\tau_{\rm dwell})$ of the electrons entering the quantum dot contributes to the shot noise. 

Following this line of argument we estimate the Fano factor $F$ (ratio of noise power and mean current) as \cite{Aga00} $F=\frac{1}{4}\exp(-\tau_{\rm E}/\tau_{\rm dwell})$, hence
\begin{equation}
F=\frac{1}{4}\left(\frac{\lambda_{F} L}{l_{\rm so}W}\right)^{1/\alpha\tau_{\rm dwell}}\;{\rm if}\;\;\frac{\lambda_{F}L}{W},\xi<l_{\rm so}<W.\label{Fsup}
\end{equation}
The upper bound on $l_{\rm so}$ indicates when diffraction at the point contact takes over as the dominant source of shot noise, while the two lower bounds indicate when full shot noise has been reached (Fano factor $1/4$) and when the softness of the boundary (penetration depth $\xi$) prevents trajectory splitting by spin-orbit interaction.

Eq.~(\ref{Fsup}) should be contrasted with the known result in the absence of spin-orbit interaction \cite{Aga00,Sil03}:
\begin{equation}\label{Fold}
F=\frac{1}{4}\left(\frac{L}{NW}\right)^{1/\alpha\tau_{\rm dwell}}\;{\rm if}\;\;\frac{\lambda_{F}L}{W}<W<l_{\rm so}.
\end{equation}
Clearly, the role of the channel number $N$ in determining the shot noise is taken over by the ratio $l_{\rm so}/\lambda_{F}$ once $l_{\rm so}$ becomes smaller than $W$.

\begin{figure}
\centerline{\includegraphics[scale=0.3]{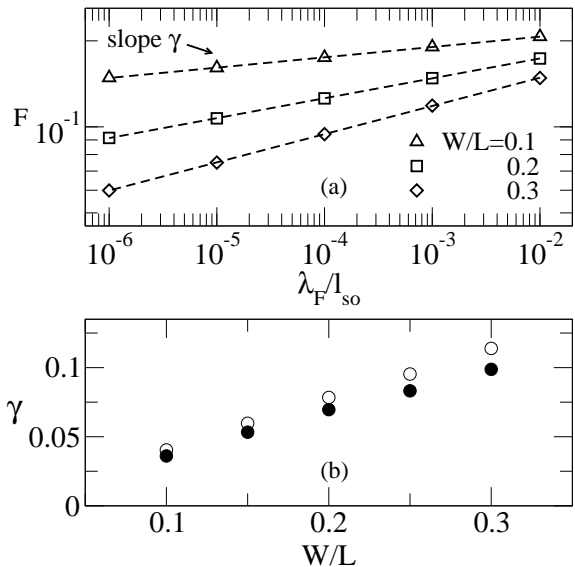}}
\caption{(a) Dependence of the Fano factor on the spin-orbit interaction strength for different widths of the opening in the billiard. The data points are calculated from Eq.~(\ref{Fano_map}). The linear fits in the log-log plot (dashed lines) confirm the predicted scaling $\log_{10} F\propto\log_{10}(\lambda_F/l_{\rm so})$. (b) Filled circles: slope 
$\gamma=d\log_{10} F/d\log_{10} (\lambda_F/l_{\rm so})$ extracted from Fig.~\ref{Fano_a}a. 
The empty circles are the theoretical prediction $\gamma=1/\alpha\tau_{\rm dwell}$.}\label{Fano_a}
\end{figure}

We support our central result (\ref{Fsup}) with computer simulations, based on the semiclassical theory of Refs.~\cite{Bla00,Pil03,Suk05}. In the limit $\lambda_{F}\rightarrow 0$ at fixed $l_{\rm so}, L, W$ a description of the electron dynamics in terms of classical trajectories is appropriate. For the spin-orbit interaction we take the Rashba Hamiltonian,
\begin{equation}
H_{\rm Rashba}=(E_{\rm so}/2p_{F})(p_{y}\sigma_{x}-p_{x}\sigma_{y}),\label{HRashba}
\end{equation}
with Pauli matrices $\sigma_{x}$ and $\sigma_{y}$. The two spin bands correspond to eigenstates of the spin component perpendicular to the direction of motion $\hat{p}$ in the $x-y$ plane (dotted arrows in Fig.\ \ref{fig_mechanism}).
The $\pm$ helicity of the spin direction $\hat{n}$ is defined by $\hat{n}\times\hat{p} = \pm \hat{z}$.
The corresponding wave vectors are
\begin{equation}
  k_\pm = \sqrt{k_F^2 + k_{\rm so}^2} \mp k_{\rm so},
\end{equation}
with $k_{\rm so} = E_{\rm so}/2v_F\hbar=\pi/l_{\rm so}$.

We consider the stadium-shaped billiard  shown in Fig.~\ref{fig_mechanism} 
with hard-wall confinement ($\xi\rightarrow 0$). Since $\lambda_F \ll L$ we can neglect the curvature of the boundary
when calculating the splitting of the trajectories by spin-orbit interaction \cite{Gov04}. The two reflection angles $\chi_\pm\in (0,\pi/2)$ of the split trajectory, measured relative to the inward pointing normal, are related by conservation 
of the momentum component parallel to the boundary,
\begin{equation}
  \label{Snell}
  k_+\sin\chi_+ = k_-\sin\chi_-.
\end{equation}
An incident trajectory of $-$ helicity is not split near grazing incidence, if $\chi_->\arcsin (k_+/k_-)\approx\pi/2-2\sqrt{k_{\rm so}/k_{\rm F}}$. Away from grazing incidence the probability $R_{\sigma\sigma^\prime}=|r_{\sigma\sigma^\prime}|^2$ for an electron incident with helicity $\sigma^\prime$ at an angle $\chi_{\sigma^\prime}$ to be reflected with helicity $\sigma$ at an angle 
$\chi_{\sigma}$ is determined by the $2\times 2$ unitary reflection matrix 
\begin{subequations}
  \label{Smatrix}
\begin{align}
  r &= \begin{pmatrix} r_{++} & r_{+-} \\ r_{-+} & r_{--} \end{pmatrix}, \\
  r_{++} &= \frac{e^{i\chi_+} - e^{-i\chi_-}}{e^{-i\chi_+}+e^{-i\chi_-}},\;\; 
  r_{--} = \frac{e^{i\chi_-} - e^{-i\chi_+}}{e^{-i\chi_+}+e^{-i\chi_-}}, \\
  r_{+-} &= -\frac{2\sqrt{\cos\chi_+\cos\chi_-}}{e^{-i\chi_+}+e^{-i\chi_-}} = r_{-+}. 
\end{align}
\end{subequations}
The reflection matrix refers to a basis of incident and reflected plane waves that carry unit flux perpendicular to the boundary, calculated using the proper spin-dependent velocity operator \cite{Mol01}.

By following the classical trajectories in the stadium billiard, and splitting them upon reflection with probabilities  $R_{\sigma\sigma^\prime}$, we calculate the probability $f(x,y,\hat{p})$ that an electron at position $x,y$ with direction $\hat{p}$ of its momentum originated from the upper left opening \cite{Bir}. The Fano factor is then given by \cite{Bla00,Pil03,Suk05}
\begin{equation}\label{Fano_map}
F=\frac{\int d\Omega\, f(1-f)}{2\int d\Omega\, f},
\end{equation}
where $d\Omega=dx\, dy\, d\hat{p}$.

\begin{figure}
\centerline{\includegraphics[scale=0.35]{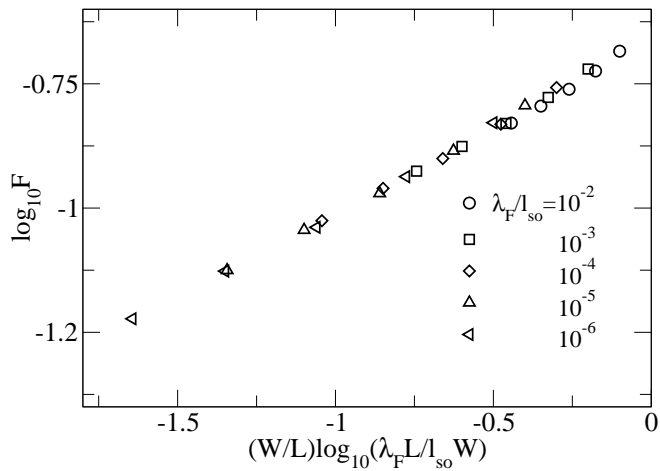}}
\caption{Dependence of the Fano factor on $W/L$ for different fixed values of  $\lambda_F/l_{\rm so}$. The data points follow closely  the predicted scaling $\log_{10} F\propto(W/L)\log_{10} (\lambda_F L/l_{\rm so}W)$.}
\label{F_wlnw}
\end{figure}

The results of the simulations are presented in Figs.~\ref{Fano_a} and \ref{F_wlnw}. We first varied $\lambda_F/l_{\rm so}$ at fixed  $W/L$ to test the scaling $F\propto(\lambda_F/l_{\rm so})^{1/\alpha\tau_{\rm dwell}}$ predicted by Eq.~(\ref{Fsup}). We kept $\lambda_F/l_{\rm so}\ll 1$, to ensure that the classical Lyapunov exponent $\alpha=0.86\,v_F/L$ \cite{Del95} and mean dwell time $\tau_{\rm dwell}$ (calculated numerically)
are not affected significantly by the spin-orbit interaction. The log-log plot in Fig.~\ref{Fano_a}a confirms the scaling   $\log_{10} F\propto\log_{10}(\lambda_F/l_{\rm so})$.  The slope $\gamma$, plotted in Fig.~\ref{Fano_a}b as a function of $W/L$ (filled circles), is close to the predicted theoretical value $\gamma=1/\alpha\tau_{\rm dwell}$ (empty circles) if the ratio $W/L$ becomes sufficiently small. 
There is no  adjustable parameter in this comparison of theory and simulation. We then tested the scaling $F\propto (L/W)^{1/\alpha\tau_{\rm dwell}}$ at fixed  $\lambda_F/l_{\rm so}$. The data points   in Fig.~\ref{F_wlnw} all fall approximately on a straight line, confirming the predicted scaling law
$\log_{10} F\propto(W/L)\log_{10}(\lambda_F L/l_{\rm so}W)$.

This completes our test of the scaling (\ref{Fsup}) in the regime $l_{\rm so}\ll W$. The scaling (\ref{Fold}), in the opposite regime $l_{\rm so}\gg W$, was verified in Ref.~\cite{Two03} using the quantum kicked rotator. We have tried to observe the crossover from the scaling (\ref{Fsup}) to (\ref{Fold}) in that model, but were not successful --- presumably because we could not reach sufficiently large system size. 
    
In conclusion, we have identified and analyzed a mechanism by which spin-orbit interaction in a ballistic system can produce electronic shot noise. The origin of the current fluctuations is a quantum mechanical effect, the splitting of trajectories, which persists in the limit of classical orbital dynamics. Since the strength of the Rashba spin-orbit interaction can be varied by a gate voltage in a two-dimensional electron gas \cite{Nit97},  the most natural way to search for the effect would be to measure the shot noise as a function of the spin-orbit precession length $l_{\rm so}$. One would then see an increase in the Fano factor  with decreasing $l_{\rm so}$, starting when $l_{\rm so}$ drops below the point contact width $W$. Since the splitting of trajectories requires $l_{\rm so}$ to be larger than the boundary penetration depth $\xi$, the noise would go down again when $l_{\rm so}$ drops below $\xi$ (assuming $\xi \ll W$). This non-monotonic dependence of the noise on the spin-orbit interaction strength would be an unambiguous signature to search for in an experiment.

This problem originated from discussions with P. W. Brouwer and V. I. Fal'ko. We have also benefited from discussions with H. Schomerus. Our research was supported by the Dutch Science Foundation NWO/FOM and by the European Community's Marie Curie Research Training Network (contract MRTN-CT-2003-504574, Fundamentals of Nanoelectronics).


\begin{thebibliography}{99}
\bibitem{Bir} {\small It is equivalent and computationally more efficient to use Birkhoff coordinates $s,p_\parallel$, with $s$ the position along the boundary and $p_\parallel$ the component of the momentum parallel to the boundary. Then 
Eq.~(\ref{Fano_map}) holds with $d\Omega=ds\, dp_\parallel$.}
\end{thebibliography}
\end{document}